\documentclass[pdflatex,sn-aps]{sn-jnl}

\pdfoutput=1
\usepackage{multirow}

\usepackage{xspace}
\newcommand{\as}{\alpha_s}
\newcommand{\zcut}{z_\text{cut}}

\newcommand{\sherpa}{S\protect\scalebox{0.8}{HERPA}\xspace}

\newcommand{\comix}{C\protect\scalebox{0.8}{OMIX}\xspace}

\newcommand{\fastjet}{F\protect\scalebox{0.8}{AST}J\protect\scalebox{0.8}{ET}\xspace}

\newcommand{\collier}{C\protect\scalebox{0.8}{OLLIER}\xspace}
\newcommand{\openloops}{O\protect\scalebox{0.8}{PEN}L\protect\scalebox{0.8}{OOPS}\xspace}

\newcommand{\jade}{J\protect\scalebox{0.8}{ADE}\xspace}

\usepackage{color}
\definecolor{darkblue}{rgb}{0,0,0.5}
\definecolor{darkgreen}{rgb}{0,0.5,0}
\definecolor{darkorange}{rgb}{0.8,0.3,0}

\newcommand\preprint[1]{\gdef\@preprint{\hfill #1}}

\begin{document}

\title{Practical Jet Flavour Through NNLO}

\author[1]{\fnm{Simone} \sur{Caletti}}\email{simone.caletti@ge.infn.it}

\author[2]{\fnm{Andrew J.} \sur{Larkoski}}\email{larkoski@slac.stanford.edu}

\author[1]{\fnm{Simone} \sur{Marzani}}\email{simone.marzani@ge.infn.it}

\author[3]{\fnm{Daniel} \sur{Reichelt}}\email{daniel.reichelt@durham.ac.uk}

\affil[1]{\orgdiv{Dipartimento di Fisica}, \orgname{Universit\`a di Genova and INFN, Sezione di Genova}, \orgaddress{\street{Via Dodecaneso 33}, \city{Genova}, \postcode{16146}, \country{Italy}}}

\affil[2]{\orgname{SLAC National Accelerator Laboratory}, \orgaddress{\street{2575 Sand Hill Road}, \city{Menlo Park}, \state{CA} \postcode{94025}, \country{USA}}}

\affil[3]{\orgdiv{Institute for Particle Physics Phenomenology, Department of Physics}, \orgname{Durham University}, \orgaddress{\city{Durham} \postcode{DH1 3LE}, \country{United Kingdom}}}

\date{\today{}}

\abstract{
An infrared and collinear (IRC) safe definition of the partonic flavour of a jet is vital for precision predictions of quantum chromodynamics at colliders.  Jet flavour definitions have been presented in the literature, but they are typically defined through modification of the jet algorithm to be sensitive to partonic flavour at every stage of the clustering.  While this does ensure that  the sum of flavours in a jet is IRC safe, a flavour-sensitive clustering procedure is difficult to apply to realistic data.  
%We introduce a distinct and novel approach to jet flavour that can be applied to any collection of partons defined by any algorithm, as long as the collection includes at least two soft emissions on the top of the hard process  \ls{referee comment 1, check}.  
We introduce a distinct and novel approach to jet flavour that can be applied to a collection of partons defined by any algorithm.
Our definition of jet flavour is the sum of flavours of all partons that remain after Soft Drop grooming, reclustered with the \jade algorithm.  
We prove that this prescription is IRC safe through next-to-next-to-leading order (NNLO), and so can interface with the most precise fixed-order calculations for jets available at present.  We validate the IRC safety of this definition with numeric fixed-order codes and further show that jet flavour with Soft Drop reclustered with a generalised $k_T$ algorithm fails to be IRC safe at NNLO.
}

\artnote{SLAC-PUB-17674\\IPPP/22/27}

\maketitle

\section{Introduction}
One of the most important and, at the same time, difficult challenges that we have to tackle in order to accurately describe fundamental particles and their interactions in high-energy collisions is posed by strong interactions. 
To date, the most effective way we have to address this issue relies on factorisation, i.e.\ on the ability of separating long-distance, non-perturbative, physics of both initial- and final-state hadrons, from the hard interaction, which can be tackled exploiting perturbative quantum field theory. 
Although very successful, this approach inevitably introduces complications. For instance, we have to map, in a quantitative way, the measurable degrees of freedom, e.g., the colliding protons or the final-state hadrons which are reconstructed by the experimental apparatuses, to the partonic degrees of freedom, i.e., quarks and gluons, that we use to describe hard interactions. 

While it is clear that quarks and gluons are not measurable degrees of freedom, many physics analyses at the CERN Large Hadron Collider (LHC), are often designed having in mind the imprint left by particular partonic flavour on a measurable final-state object, such as a hadron or a jet. 
Furthermore, the issue of jet flavour acquires particular relevance when discussing heavy quarks. In this case indeed one can meaningfully assign a flavour-label to a jet exploiting kinematic features of $D$ and $B$ meson decays, such as, for instance, displaced vertices. 

Naively one would be tempted to call jet flavour the net flavour of the jet after the generalised $k_T$-clustering \cite{Catani:1993hr,Ellis:1993tq,Dokshitzer:1997in,Wobisch:1998wt,Wobisch:2000dk,Cacciari:2008gp}, i.e., simply to compute the total number of quarks minus anti-quarks for each quark flavour. However, this procedure is not infrared and collinear (IRC) safe at next-to-next-to leading order (NNLO), as pointed out in~\cite{Banfi:2006hf} (BSZ).
The problematic configuration is depicted in Fig.~\ref{fig:sd_nnlo_fail}, where we show an $\mathcal{O}(\alpha_s^2)$ configuration characterised by the emission of a soft gluon, which splits into a quark--anti-quark pair, $q\bar q$. In this configuration, the jet algorithm clusters together the hard quark $Q$ and $q$ and so the jet flavour is determined by the soft gluon splitting, rendering the flavour assignment IRC unsafe. 
BSZ solved this problem by modifying the metric of the clustering algorithm so that clustering of soft pairs are favoured only if the softer parton is flavoured. 
This so-called flavour-$k_T$ algorithm has been used in precision calculations~\cite{Gauld:2020deh,Czakon:2020coa} (see also \cite{Catani:2020kkl}).

However, the use of the flavour-$k_T$ algorithm in experimental analysis is far from straightforward. The main complication arises from the fact that the clustering metric requires knowledge of the flavour of the objects combined at every stage of the clustering. This clashes with the experimental procedure of assigning the flavour label to the jet after clustering, rather than to the jets' constituents. Furthermore, the LHC experimental collaborations have put a lot of effort in standard jet calibration and may be somewhat reluctant to dramatically change their jet definition strategies. Despite the fact that some practical solutions to these problems have been suggested in~\cite{Banfi:2007gu}, nowadays comparison to high-precision QCD calculations with LHC data typically involve unfolding corrections that bridge the gap between the theoretical and experimental jet definitions, see for instance~\cite{Gauld:2020deh} for the case of $b$-jets. These corrections are typically derived using Monte Carlo simulations and, while in some cases turn out to be modest, they put us in an uneasy situation because they degrade the theoretical accuracy of the calculation. 

In this study we suggest an alternative approach to jet flavour assignment, which is IRC safe through NNLO, while maintaining experimental viability.  Our suggestion stems from the realisation that the problematic configuration depicted in Fig.~\ref{fig:sd_nnlo_fail} is analogous to the same configuration responsible for non-global logarithms~\cite{Dasgupta:2001sh} in jet shape observables. In fact, similar to non-global logarithms, it is possible to eliminate the infrared ambiguities of jet flavour with Soft Drop (SD) grooming \cite{Larkoski:2014wba}.  There are, however, two restrictions or modifications to SD that we must implement to ensure IRC safety of jet flavour through NNLO.  First, the angular exponent $\beta$ in the grooming constraint must be greater than 0 for IRC safety of jet flavour at next-to-leading order (NLO).  This restriction eliminates the modified Mass Drop Tagging Groomer (mMDT) \cite{Dasgupta:2013ihk,Dasgupta:2013via} for use in defining IRC safe flavour.  It is also known that the mMDT jet energy or transverse momentum to the beam is also not IRC safe, for similar reasons \cite{Larkoski:2014wba,Baron:2018nfz}.

Secondly, the \jade clustering algorithm \cite{JADE:1986kta,JADE:1988xlj} must be used with SD to properly order and groom soft emissions that can render the jet flavour ambiguous.  SD or mMDT was originally introduced to groom emissions in a jet ordered in relative angle, with the Cambridge/Aachen algorithm \cite{Dokshitzer:1997in,Wobisch:1998wt,Wobisch:2000dk}.  When typical IRC safe observables are measured, like the mass, any generalised $k_T$ algorithm can be used to order emissions in the jet and then groom with SD and produce IRC safe results upon grooming.  However, we will explicitly demonstrate that there exist orderings of partons in a jet at NNLO with generalised $k_T$ whose flavour cannot be made IRC safe through SD grooming.  \jade cures this problem by always clustering soft quarks together first, which then can be eliminated from the jet by SD.  We note that it has been observed that jet flavour is IRC safe through NNLO for jets clustered with the \jade algorithm \cite{Banfi:2006hf}.  However, the way that \jade is used here is distinct: a jet is defined by whatever algorithm the user desires.  Then, the emissions in the identified jet are reclustered with \jade and SD groomed.  We prove that the jet flavour defined by the sum of particles that remain after this grooming procedure is IRC safe through NNLO.  In a companion paper \cite{Caletti:2022glq}, we exploit perturbative fragmentation functions to introduce a definition of jet flavour that is soft safe, but not collinear safe, for jets with arbitrary numbers of constituents.

In this paper, we explicitly consider jet production in $e^+e^-$ collisions. However, our analytic calculations are performed in the collinear limit, which is universal. Therefore, our results can be applied to hadron-hadron collisions, with appropriate changes of detector coordinates.

The outline of this paper is as follows. In Sec.~\ref{sec:sd_flavour}, we study jet flavour after grooming with SD and generalised $k_T$ reclustering at NLO and NNLO and show that IRC safety fails at NNLO.  In Sec.~\ref{sec:sd_flavour_jade}, we modify the SD groomer with \jade reclustering, and argue that the flavour defined by this algorithm is IRC safe through NNLO.  Sec. \ref{sec:numerical} reports fixed-order numerical calculations obtained using \sherpa \cite{Sherpa:2019gpd}, validating the IRC safety of the flavour algorithm. We conclude in Sec.~\ref{sec:conclusions}.  Explicit definitions of the SD grooming algorithm, the Durham $k_T$ jet algorithm \cite{Catani:1991hj}, and the BSZ flavour-$k_T$ algorithm are presented in appendices.

\section{Jet Flavour from Grooming}\label{sec:sd_flavour}
We start our discussion about SD flavour by considering the leading-order (LO) and NLO situations first. Through NLO, even standard jet algorithms provide an IRC safe definition of jet flavour, but it is still interesting to work through the calculation in order to understand the role of grooming. Let us consider, for definiteness, a jet initiated by a quark. For simplicity, we  will assume that the jet radius $R$ is small, so we can work in the collinear limit.  We will  calculate the jet flavour as defined by the sum of partonic flavour of the more energetic jet.

At LO, relative ${\cal O}(\alpha_s^0)$, there is only one particle in the jet, the initiating quark, which trivially passes the SD condition and so the jet has quark flavour.  Thus, there is 0 probability that the jet to this order is gluon flavour and the jet flavour fractions to this order are
\begin{align}
&P_q = 1+{\cal O}(\alpha_s)\,,
&P_g = 0+{\cal O}(\alpha_s)\,.
\end{align}

\subsection{Soft Drop Flavour at NLO}\label{subsec:sd_flavour_nlo}
At NLO, relative ${\cal O}(\alpha_s)$, the quark can emit a real gluon through
$q \to q g$ splitting. It is easier to first determine the configurations that
lead to assigning gluon flavour to the jets. This can happen for two reasons:
(a) either the quark and the gluon are not recombined in the same jet (i.e., the
angle $\theta$ between them is bigger than $R$) and the gluon jet is more
energetic (i.e., the gluon momentum fraction $z>\frac{1}{2}$) or (b) the two partons are recombined in the same jet but the quark fails the SD condition and it is groomed away.  In the former case we have 
\begin{align}
 P_g^{(a)}&=\frac{\alpha_s C_F}{2\pi} \int_{R^2}^1\frac{d\theta^2}{\theta^{2}} \int_{1/2}^1 dz\, 
\frac{1+(1-z)^2}{z}
\\
&
=\frac{\alpha_s C_F}{2\pi}\, \log R^2 \left(
\frac{5}{8}-2\log 2
\right)\,.\nonumber
\end{align}
In the evaluation of this result, we only keep the leading terms in the $R^2\ll 1$ limit.

For the latter case, the quark and the gluon live in the same jet but the quark is groomed away by SD. A review of the SD grooming algorithm is presented in App.~\ref{app:sdalg}. To fail SD, the quark's energy fraction $1-z$ and its splitting angle from the gluon $\theta$ must satisfy
\begin{align}
\zcut\left(\frac{\theta^2}{R^2}
\right)^\beta>1-z\,,
\end{align}
where we have assumed that $\theta,R\ll 1$.  Then, the probability that the quark is groomed away is
\begin{align}
 P_g^{(b)}&=\frac{\alpha_s C_F}{2\pi}\int_0^{R^2}\frac{d\theta^2}{\theta^{2}} \int_0^1dz\, 
\frac{1+(1-z)^2}{z}\,\Theta\left(\zcut\left(
\frac{\theta^2}{R^2}
\right)^\beta-(1-z)\right)\\
&
=\frac{\alpha_s C_F}{2\pi}\frac{\zcut}{\beta}
\,,\nonumber
\end{align}
to leading order in $\zcut \ll 1$.

In total, the gluon and quark flavour fractions defined by SD grooming through ${\cal O}(\alpha_s)$ are:
\begin{align}
P_g &= P_g^{(a)}+ P_g^{(b)}=\frac{\alpha_s C_F}{2\pi}\, \left( \left(
\frac{5}{8}-2\log 2
\right)\log R^2 + \frac{\zcut}{\beta}
\right)+{\cal O}(\alpha_s^2)\,, \nonumber \\
P_q &= 1-P_g= 1+\frac{\alpha_s C_F}{2\pi}\left(
\left(
2\log 2 - \frac{5}{8}
\right)\log R^2-\frac{\zcut}{\beta}
\right)+{\cal O}(\alpha_s^2)\, .
\end{align}
This is finite for $\beta > 0$, but is not for $\beta = 0$.  Thus, jet flavour, even at ${\cal O}(\alpha_s)$, is only IRC safe for SD with $\beta  > 0$. This is not a surprise, because when used as a groomer, SD with $\beta=0$, i.e., the mMDT groomer, is only IRC safe when the substructure of the jet is resolved. 

\subsection{Soft Drop Flavour at NNLO}\label{subsec:sd_flavour_nnlo}
At $\mathcal{O}(\alpha_S^2)$ there are several configurations of particles that must be considered. In fact, both $q\rightarrow qg\rightarrow qgg$ and $q\rightarrow qg\rightarrow q\,q'\,\bar q\,'$ splittings contribute, where in principle $q'\,(\bar q\,')$ is a (anti)quark with a different flavour with respect to the initial quark $q$. The problematic configuration, pointed out by BSZ and represented by Fig.~\ref{fig:sd_nnlo_fail}, belongs to the second case. In this picture we labeled the hard initial quark as $Q$ while the soft quark--anti-quark pair has been labeled as $q\bar{q}$, %because of the natural application of the algorithm to heavy-quark ($Q$) flavoured jets. 
and we keep this naming convention in what follows.
The configuration represented in Fig.~\ref{fig:sd_nnlo_fail} spoils naive IRC safety of jet flavour because only one of the soft quark--anti-quark pair emitted by the soft gluon will be in the partonic content of the final jet varying its net flavour. In this section, we will discuss how default SD cures this issue, but still fails to be IRC safe through NNLO when emissions in the jet are ordered with a generalized $k_T$ algorithm. This will motivate the introduction of a modification to SD in the next section, in which emissions are reclustered with the \jade algorithm, and we will prove that it results in IRC safe jet flavour through NNLO.

\subsubsection{Elimination of Soft Quark Ambiguities}

\begin{figure}
\begin{center}
\includegraphics[width=3cm]{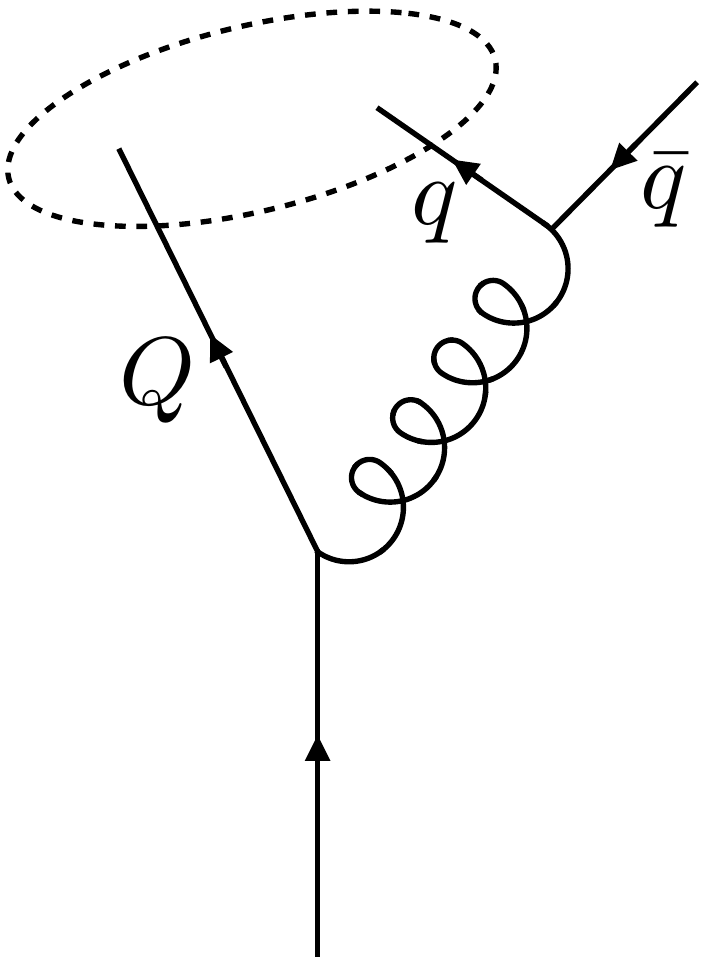}
\caption{\label{fig:sd_nnlo_fail}
The configuration that renders jet flavour definition infrared unsafe at NNLO is depicted: a quark $Q$ emits an intermediate soft gluon that subsequently splits into a quark--anti-quark $q\bar q$ pair. Only one of the gluon's decay products, say $q$, is clustered with the original quark $Q$ and so the jet flavour is determined by soft physics. Note that the dotted oval can either represent the boundary of the original jet or the effective boundary induced by SD.
}
\end{center}
\end{figure}

The configuration in Fig.~\ref{fig:sd_nnlo_fail} in which the dashed oval
represents the jet boundary is essentially the same configuration of particles
that are the leading contribution to non-global logarithms (NGLs)
\cite{Dasgupta:2001sh}.  Though at NNLO, the jet consists of only two particles,
and so the implementation of SD on the jet is identical to that at NLO.  The
softer of the two constituents of the jet is eliminated by the groomer if it
fails the SD constraint.  With a finite value of $\zcut$ and $\beta$, an
arbitrarily soft quark $q$ will always fail the SD constraint, and so after
grooming the jet will consist exclusively of the hard quark $Q$.
  Thus, in the soft limit, the jet flavour would be identified as the same flavour as $Q$, which is also the flavour of the jet from corresponding virtual corrections.  Thus, this configuration has no infrared ambiguities. ~\footnote{Kinematically, the two quarks $Q$ and $q$ can become collinear, thus passing the SD condition. However, no collinear singularity is associated with this configuration.}

Further, because of the relationship to NGLs, all-orders statements about the jet flavour from this configuration can be made.  It has been proven that SD and mMDT grooming eliminate NGLs of observables like the jet mass to all orders in perturbation theory \cite{Dasgupta:2013ihk,Larkoski:2014wba,Frye:2016aiz}.  NGLs arise from soft particles that are sensitive to the boundary of the jet.  Correspondingly, the jet flavour as defined by application of SD has no infrared divergences arising from soft emissions near the boundary of the jet.  By contrast, SD is inclusive over collinear emissions at the jet center, and we will demonstrate that this feature is problematic for jet flavour.

\subsubsection{Failure of IRC Safety of SD with $k_T$ Clustering}\label{sec:nnloircunsafe}

In the original and most widely-studied definitions of SD grooming, emissions in the jet are re-clustered with a generalised $k_T$ algorithm, typically the Cambridge/Aachen (C/A) algorithm \cite{Dokshitzer:1997in,Wobisch:1998wt,Wobisch:2000dk} in which emissions are ordered by their relative angle.  While this prescription does eliminate the NGL-like infrared ambiguities in jet flavour, reclustering with a $k_T$-like algorithm means that the emission that first passes the groomer sets an effective jet radius below which all emissions are still included in the jet.  Thus, a configuration as illustrated in Fig.~\ref{fig:sd_nnlo_fail} can still exist, where now the dashed oval represents the effective groomed jet region.  That is, grooming can eliminate a soft, wider-angle anti-quark from the jet, but render the jet flavour ambiguous because a soft quark passes the groomer.  In this section, we will make this precise, and explicitly demonstrate that a default implementation of the SD groomer still fails to give an IRC safe jet flavour at NNLO.

For simplicity of expressions, we will restrict our analysis here to consideration of C/A clustering of emissions in the jet.  Our jet of interest will initially consist of a hard quark $Q$ and a soft quark--anti-quark pair $q\bar q$ from intermediate gluon emission.  Then, on this collection of particles we will groom with SD, necessarily assuming that the angular exponent $\beta > 0$ to ensure IRC safety at NLO.  Then, the SD constraint represented by Fig.~\ref{fig:sd_nnlo_fail} in which the anti-quark is at a wider angle than the quark to $Q$ and the anti-quark fails the groomer while the quark passes is
\begin{align}
&\Theta_\text{SD}^\text{C/A} \\
&= \Theta(\theta^2_{Q\bar q}-\theta^2_{Qq})\Theta(\theta^2_{q\bar q}-\theta^2_{Qq})\Theta\left(z_q-\zcut \left(\frac{\theta_{Qq}^2}{R^2}\right)^\beta\right)\Theta\left(\zcut \left(\frac{\theta_{Q\bar q}^2}{R^2}\right)^\beta-z_{\bar q}\right)\,,\nonumber
\end{align}
where we assume that the soft quark and anti-quark energy fractions $z_q,z_{\bar q}\ll 1$.  Pairwise angles between particles are labeled; i.e., $\theta_{Qq}$ is the angle between the hard quark $Q$ and the soft quark $q$.  The first two $\Theta$ functions are the implementation of the C/A clustering, while the latter two $\Theta$ functions are the SD groomer constraints on the soft quark and anti-quark.

To isolate the problematic, IRC-unsafe configuration, we can rescale the energy fractions in the collinear limit as
\begin{align}\label{eq:zqqbar_rescale}
&z_q = x_q \zcut \left(\frac{\theta_{Qq}^2}{R^2}\right)^\beta\,,&z_{\bar q}=  x_{\bar q} \zcut \left(\frac{\theta_{Q\bar q}^2}{R^2}\right)^\beta\,,
\end{align}
for some new quantities $x_q,x_{\bar q}$.  Assuming that the ratio of the angles
with respect to the hard quark stays constant in approaching the collinear
limit, $\theta_{Qq}\lesssim \theta_{Q\bar q}\ll R$, this change of variables
exposes the collinear singularity. In this limit, the matrix element $\vert
  {\cal M}(z_q,z_{\bar q})\vert ^2$ can be thought of as essentially the triple
  collinear splitting function, for which we give an explicit expression in
  App.~\ref{app:tricoll}. Under the above rescaling the matrix element
 and differential phase space measure
$d\Pi_3$ are only modified by an order-$1$ amount set by the ratio
$\theta_{Qq}/\theta_{Q\bar q}$ in the soft and collinear limit: 
\begin{align}
&d\Pi_3\, \vert {\cal M}(z_q,z_{\bar q})\vert ^2\, \Theta_\text{SD}\\
&
\hspace{1cm}\simeq d\Pi_3\, \vert {\cal M}(x_q,x_{\bar q})\vert ^2\, \Theta(\theta^2_{Q\bar q}-\theta^2_{Qq})\Theta(\theta^2_{q\bar q}-\theta^2_{Qq})\Theta\left(x_q-1\right)\Theta\left(1-x_{\bar q}\right)\,.\nonumber
\end{align}
Here, $\simeq$ means equal up to order-$1$ factors. This change of variables
decouples the energy fractions and splitting angles to leading power, and
exposes the collinear divergence of the matrix element, rendering SD flavour
with the Cambridge/Aachen algorithm IRC unsafe at NNLO. This IRC unsafe argument
extends to SD with general $k_T$ reclustering because the constraint on the
orderings of the branches is homogeneous in the energy fraction of the soft
quarks and so the rescaling of Eq.~\ref{eq:zqqbar_rescale} does not dominantly
change the branching structure. 

\section{Soft Drop Flavour with \jade Reclustering}\label{sec:sd_flavour_jade}

The key issue with the IRC safety of SD flavour was due to the features of $k_T$ reclustering.  As observed in \cite{Banfi:2006hf}, the $k_T$ class of algorithms does not favor clustering two soft particles together first, if there is a hard particle around at smaller angle.  However, if the soft quark--anti-quark pair were clustered together first, then SD would simply groom them away, which would produce no effect on the jet flavour as simply defined from the hard quark $Q$.  Therefore, we will modify the SD grooming procedure to ensure that the softest pair of particles is clustered first.  This can be accomplished through NNLO with the \jade algorithm \cite{JADE:1986kta,JADE:1988xlj}.

Our procedure for achieving an IRC safe definition of jet flavour through at least NNLO accuracy of an arbitrary collection of particles in a pre-defined jet is as follows.  We express the procedure in phase space coordinates appropriate for jets produced in $e^+e^-$ collisions and for jets in hadron collisions, one exchanges energies for momentum transverse to the beam and angles for longitudinal boost-invariant angles.
\begin{enumerate}

\item Recluster the jet with the \jade algorithm which has a metric $d_{ij}$ corresponding to the pairwise mass of particles:
\begin{align}
d_{ij}^\text{\jade} = 2E_i E_j(1-\cos\theta_{ij})\,.
\end{align}

\item At each stage of the clustering, require that particles $i$ and $j$ pass the SD grooming requirement, where:
\begin{align}
\frac{\min[E_i,E_j]}{E_i+E_j} > \zcut \left(
\frac{\theta_{ij}^2}{R^2}
\right)^\beta\,,
\end{align}
with the initial jet radius $R$, angular exponent $\beta > 0$, and energy scale parameter $0 < \zcut < 1/2$.

\item If the stage in the clustering passes the grooming requirement, terminate and return the sum of flavours of particles in the jet.  If the grooming requirement fails, then remove the softer of the two branches, and continue to the next stage of the \jade clustering along the harder branch.
\end{enumerate}

\subsection{Argument for IRC Safety Through NNLO}\label{sec:safe_nnlo}

With this new flavour algorithm, we return to the configuration of Fig.~\ref{fig:sd_nnlo_fail} and explicitly show that its contribution to the jet flavour is IRC safe.  With \jade clustering for SD, this problematic configuration has phase space constraints of the form:
\begin{align}
\Theta_\text{SD}^\text{JADE} &= \Theta(m^2_{Q\bar q}-m^2_{Qq}) \Theta(m^2_{q\bar q}-m^2_{Qq})
\\ &
\hspace{1cm} \times
\Theta\left(z_q-\zcut \left(\frac{\theta_{Qq}^2}{R^2}\right)^\beta\right)\Theta\left(\zcut \left(\frac{\theta_{Q\bar q}^2}{R^2}\right)^\beta-z_{\bar q}\right)\,,\nonumber
\end{align}
where now pairwise particle invariant masses are compared in the first two $\Theta$ functions.  Under the same change of variables as Eq.~\ref{eq:zqqbar_rescale}, the mass orderings take a different form where
\begin{align}
\Theta(m^2_{Q\bar q}-m^2_{Qq}) &= \Theta\left(z_Qx_{\bar q} \zcut \left(\frac{\theta_{Q\bar q}^2}{R^2}\right)^\beta\theta^2_{Q\bar q}-z_Qx_{q} \zcut \left(\frac{\theta_{Q q}^2}{R^2}\right)^\beta\theta^2_{Qq}\right)\\
&=\Theta\left(x_{\bar q} \theta_{Q\bar q}^{2(\beta+1)}-x_{q} \theta_{Qq}^{2(\beta+1)}\right)
\nonumber\,,\\
\Theta(m^2_{q\bar q}-m^2_{Qq}) &= \Theta\left(x_q \zcut \left(\frac{\theta_{Q q}^2}{R^2}\right)^\beta  x_{\bar q} \zcut \left(\frac{\theta_{Q\bar q}^2}{R^2}\right)^\beta\theta^2_{q\bar q} 
\right.  \nonumber\\ & \quad \quad \quad \quad \quad \quad \quad \quad \quad \quad \quad \quad \quad \quad    \left.
-z_Qx_{q} \zcut \left(\frac{\theta_{Q q}^2}{R^2}\right)^\beta\theta^2_{Qq}\right)\nonumber\\
&=\Theta\left( x_{\bar q} \zcut \left(\frac{\theta_{Q\bar q}^2}{R^2}\right)^\beta\theta^2_{q\bar q}-\theta^2_{Qq}\right)
\,.
\end{align}
In writing these expressions, we are working in the collinear limit for all pairwise masses and assume that the hard quark $Q$ takes (nearly) all of the energy, $z_Q \to 1$.  In these coordinates, the SD constraint with \jade reclustering becomes:
\begin{align}
\Theta_\text{SD}^\text{JADE} &=\Theta\left(x_{\bar q} \theta_{Q\bar q}^{2(\beta+1)}-x_{q} \theta_{Qq}^{2(\beta+1)}\right)\Theta\left( x_{\bar q} \zcut \left(\frac{\theta_{Q\bar q}^2}{R^2}\right)^\beta\theta^2_{q\bar q}-\theta^2_{Qq}\right)\\
&
\hspace{2cm}\times\Theta\left(x_q-1\right)\Theta\left(1-x_{\bar q}\right)\,.\nonumber
\end{align}

Now we see immediately that the ordering of emissions that \jade imposes regulates the divergent regions.  For example, if the anti-quark $\bar q$ that fails SD becomes arbitrarily soft, $x_{\bar q}\to 0$, the constraint that $m_{q\bar q}^2 > m_{Qq}^2$ fails.  Instead, we could take the collinear limit, where all angles $\theta^2\to 0$ at a similar rate.  However, with $\beta > 0$, we observe again that the constraint $m_{q\bar q}^2 > m_{Qq}^2$ fails.  Finally, we can consider a correlated soft/collinear limit such that the constraint $m_{q\bar q}^2 > m_{Qq}^2$ is satisfied.  We can isolate this limit by introducing the scaling parameter $\lambda>0$ and require that
\begin{align}
&x_{\bar q}\to \lambda x_{\bar q}\,, &\theta^2 \to \lambda^{-\frac{1}{\beta}} \theta^2\,,
\end{align}
for any pairwise angle $\theta^2$.  This scaling preserves the constraint that 
\begin{align}
x_{\bar q} \zcut \left(\frac{\theta_{Q\bar q}^2}{R^2}\right)^\beta\theta^2_{q\bar q}>\theta^2_{Qq}\,,
\end{align}
by construction.  However, the constraint that $m^2_{Q\bar q}>m^2_{Qq}$ is rescaled to
\begin{align}
x_{\bar q} \theta_{Q\bar q}^{2(\beta+1)}>x_{q} \theta_{Qq}^{2(\beta+1)}\qquad \to\qquad \lambda x_{\bar q} \theta_{Q\bar q}^{2(\beta+1)}>x_{q} \theta_{Qq}^{2(\beta+1)}\,,
\end{align}
which is clearly violated for sufficiently small $\lambda$.  Therefore, the jet flavour defined as the sum of flavours that remain in a jet after SD with \jade reclustering is IRC safe, through NNLO.

\begin{figure}
\begin{center}
\includegraphics[width=3.2cm]{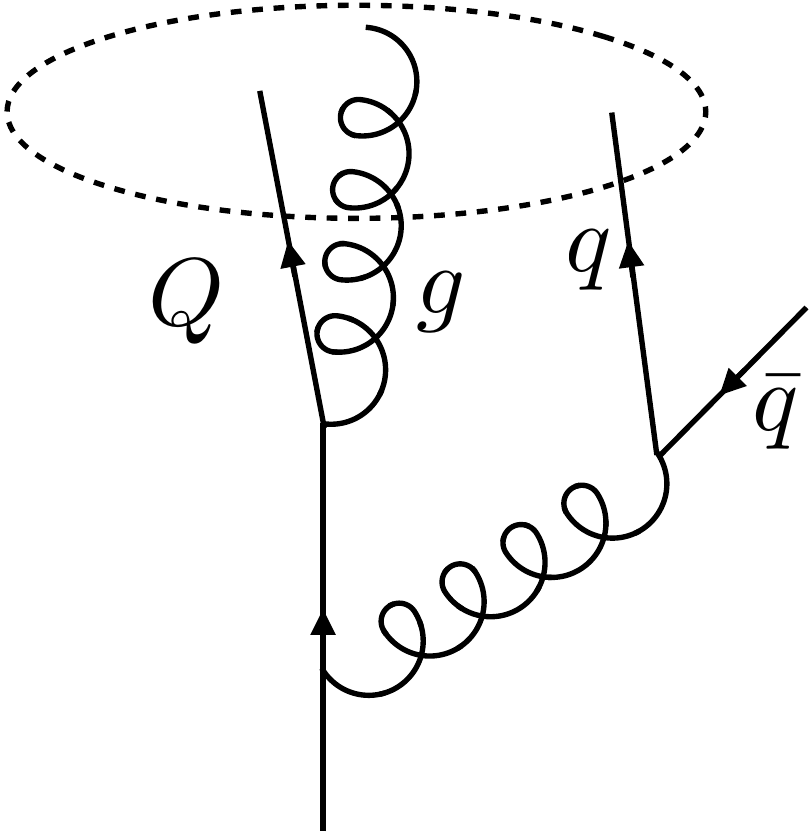}
\caption{\label{fig:nnnlo_fail}
Illustration of a configuration of particles that renders SD flavour with \jade reclustering IRC unsafe at NNNLO.  The jet boundary is illustrated in the dashed oval, with the hard quark $Q$ and hard gluon $g$ with the largest pairwise mass and pass SD.  An arbitrarily soft quark $q$ lands in the jet and is therefore never groomed away, rendering the jet flavour ambiguous.
}
\end{center}
\end{figure}

However, we do not expect this jet flavour definition to be IRC safe at higher perturbative orders.  We illustrate one configuration at next-to-next-to-next-to-leading order (NNNLO) in Fig.~\ref{fig:nnnlo_fail} that demonstrates the problem.  The jet boundary is illustrated by the dashed oval, and the particles in the jet consist of a hard quark $Q$, a hard gluon $g$, and a soft quark $q$.  The partner soft anti-quark $\bar q$ is not clustered into the jet.  We assume that the hard quark and gluon are sufficiently collinear and have the largest pairwise mass, and are therefore de-clustered first with \jade.  With $\beta > 0$, collinear particles always pass SD, and so the soft quark $q$ is not groomed and necessarily remains in the jet.  This remains true for an arbitrarily low energy of the quark, and so this definition of jet flavour will not be IRC safe at NNNLO.

However, several modifications to SD have been proposed that may solve
this IRC unsafety issue.  In particular, techniques that continue to apply soft
drop after the first emission passes, e.g.,
Refs.~\cite{Frye:2017yrw,Dreyer:2018tjj,Mehtar-Tani:2019rrk}, may eliminate
flavour ambiguities at higher orders when combined with \jade reclustering.  We
leave a detailed study of this possibility and the necessary features of such a
groomer to future work. 

\section{Numerical Results}\label{sec:numerical}

We now perform numerical tests to validate the IRC safety of jet flavour from SD with \jade reclustering. 
We consider jet production in $e^+ e^-$ collisions, with jets defined using
the Durham clustering algorithm~\cite{Catani:1991hj} with resolution parameter
$y_\text{cut}$, see App.~\ref{app:durham} for details.  Two exclusive jets are found, and we determine the flavour in each jet separately.
We will perform the tests with SD parameters $\beta=2$ and $\zcut=0.1$.
Following BSZ, we now introduce the $3$-jet resolution parameter $y_3$,
i.e.\ the maximum value of $y_\text{cut}$ for which the event has $3$ jets.
We perform the calculations within the \sherpa \cite{Sherpa:2019gpd}
framework. A version of the BSZ algorithm for flavour identification has been
implemented and tested in \cite{Baberuxki:2019ifp, Baron:2020xoi, Caletti:2021oor}. We modified
this to include our flavour definition, based on the \fastjet \cite{Cacciari:2011ma}
implementation of SD grooming. The fixed order matrix elements are
calculated using \comix \cite{Gleisberg:2008fv}, and one loop virtual
corrections are obtained from \openloops \cite{Cascioli:2011va}, relying on the
\collier \cite{Denner:2016kdg}  library. Infrared divergences between the two
are regularised using the Catani-Seymour subtraction method \cite{Catani:1996vz,
  Gleisberg:2007md}. For concreteness, we perform all calculations at a center
of mass energy corresponding approximately to the Z pole, $\sqrt{s} =
91.2~\text{GeV}$, and set the renormalisation scale $\mu_R$ to that value.

We can use $y_3$ as a slicing parameter and write the inclusive NNLO cross section as the sum of two contributions, above and below the cut:
\begin{equation}\label{eq:nnlo_slice}
\sigma^\text{NNLO}=  \int_0^{y_3}  d y_3'\,  \frac{d \sigma}{d y_{3}'}+\int_{y_3}^{y_\text{max}} d y_3' \, \frac{d \sigma}{d y_{3}'},
\end{equation}
where in the first contribution we would have to include the 2-loop virtual corrections, while the second has an extra emission and so can be evaluated at 1-loop. 
In order to establish IRC safety it is enough to study the behaviour of the NLO distribution $\frac{d \sigma}{d y_3}$ at small $y_3$. We will find logarithmic divergences, but if IRC safety holds these are all cancelled by the below-the-cut contribution, i.e.\ the first term in Eq.~\ref{eq:nnlo_slice}.
In order for this cancellation to take place, the flavour assignment of the two contributions must coincide in the singular $y_3\to 0$ limit.  In turn, this implies that all flavour assignments that do not exist at Born level give a vanishing contribution to $y_3 \frac{d \sigma}{d y_3}$, in the $y_3 \to 0$ limit. 

\begin{figure}
\begin{center}
  \includegraphics[width=.49\textwidth]{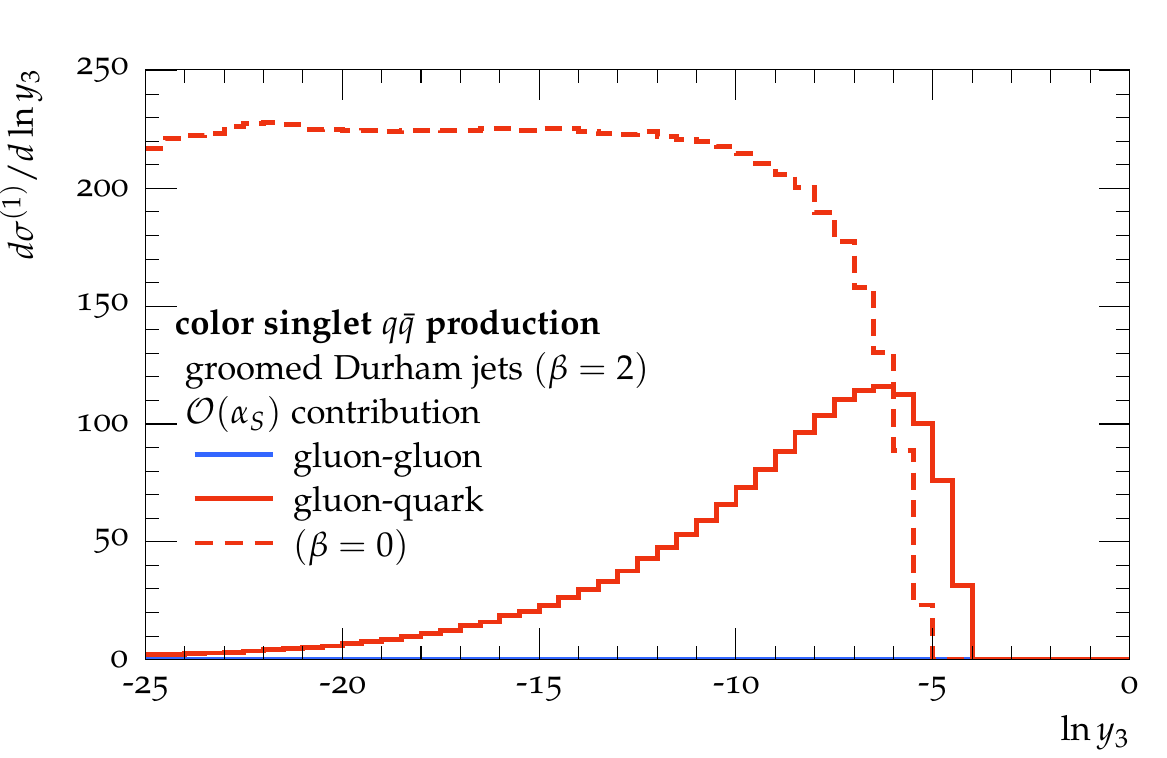}
  \includegraphics[width=.49\textwidth]{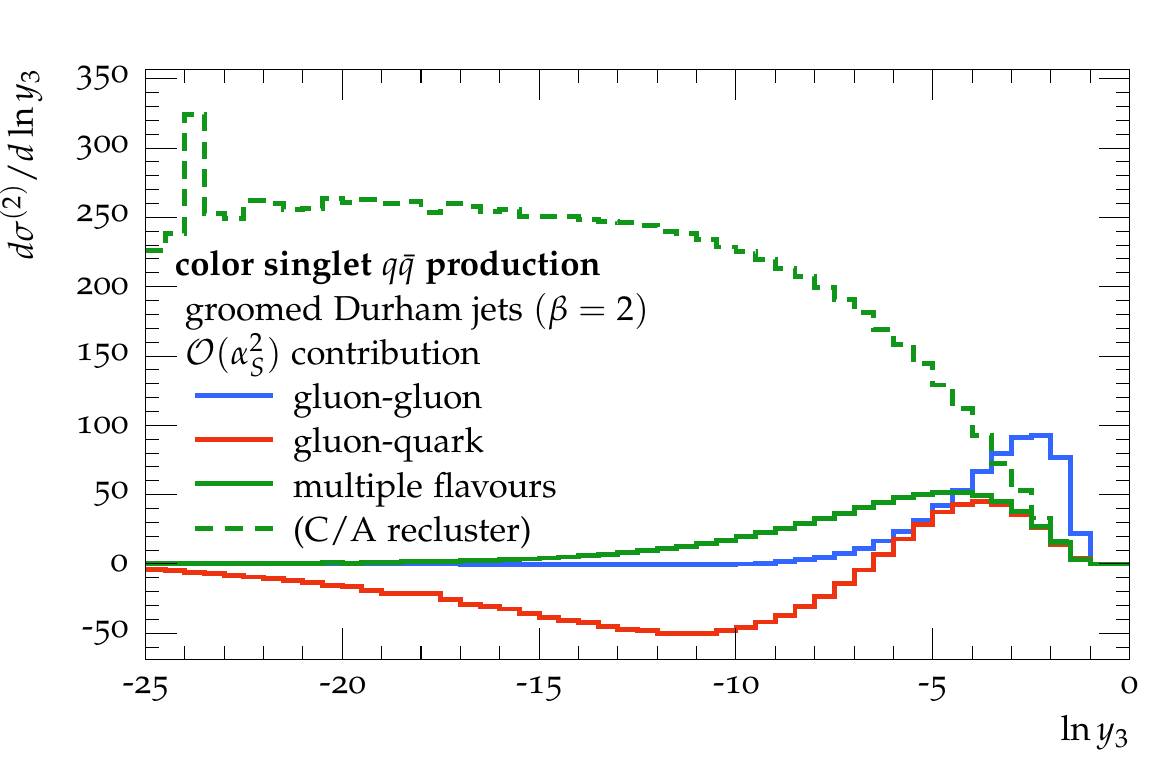}
\caption{NLO (left) and NNLO (right) contributions to the cross section as a function of the $y_3$ jet
  resolution of the event, for different assignments of flavour to the two jets
  obtained from Durham clustering, according to the jet constituents after SD grooming with $\beta = 2$ reclustered with \jade (solid).  Two IRC-unsafe flavour definitions are also shown dashed: on left, jet flavour with $\beta = 0$ SD/mMDT grooming and on right, $\beta = 2$ SD grooming but with C/A reclustering.}\label{numerical_tests}
\end{center}
\end{figure}

Let us first perform the test for the $\mathcal{O}(\as)$ contribution,
i.e., the lowest order real correction to inclusive jet production. This is shown
in the left of Fig.~\ref{numerical_tests}. As in the case of the flavour of
either plain Durham jets or BSZ jets, there are hard configurations where the
quark and anti-quark are clustered together, and hence the event is identified as having
two gluon flavour jets. This contribution vanishes for any flavour definition as $y_3\to 0$ because there are no singularities associated to clustering the hard quark--anti-quark together.
In addition, since SD can remove flavoured objects from the
event, we now also have contributions where one jet is identified as gluon and
the other as a quark jet. It also vanishes as $y_3\to 0$ for finite
$\beta$. However, for $\beta=0$ it does not, which we additionally illustrate
explicitly, consistent with the analysis of Sec.~\ref{subsec:sd_flavour_nlo}. 

Next, we can perform the numerical test for the $\mathcal{O}(\as^2)$
contribution. The result is shown in the right hand plot in
Fig.~\ref{numerical_tests}. The same configurations as at the first order
appear. In addition, with the higher multiplicity it is now possible that jets contain multiple flavours. This includes quark--anti-quark pairs of
different flavours, as well as multiple quarks without matching
anti-quarks. There are configurations where both jets are like this, or one of
them could still be identified as a quark. All of them have to vanish, so we
here collect them into one common contribution. We again illustrate that this
test fails for the last contribution if the jets are reclustered with C/A
instead of \jade.

We have performed the same tests for colour singlet gluon production from Higgs
decays. The explicit results can be found in App.~\ref{app:num_gg_final}. They
confirm that all channels, apart from the gluon-gluon one in  this case, vanish
in the soft limit. This numerically validates the results of the previous
sections for all possible configurations at NNLO.

\section{Conclusions}\label{sec:conclusions}

We introduced a novel definition of jet flavour that is IRC safe through NNLO.  No modification to the jet clustering in the event is required, and properties of jet grooming are exploited to ensure IRC safety.  However, to ensure that soft quark pairs that can render jet flavour ambiguous are always clustered together and then groomed away, we must recluster the jet with the \jade algorithm.  Typically, the SD groomer is reclustered with a $k_T$-class algorithm, which is vital for calculability and factorization of observables measured on the groomed jet.  The \jade algorithm is known to violate soft-collinear factorization, but here we demonstrate that this flaw is actually a necessary feature for jet flavour that is insensitive to partonic flavour during reclustering.

Our jet flavour definition is limited to NNLO accuracy, which is sufficient for implementation with the highest fixed-order predictions available at present.  However, one would desire an all-orders, IRC safe, flavour definition that could be implemented on a jet with an arbitrary number of constituents.  Such a jet flavour definition could then potentially be calculated in resummed perturbation theory or for which evolution equations could be derived.  We have argued that naive extensions of the algorithm presented here fail at NNNLO and further, the use of \jade reclustering may present obstructions for resummation.  Iterative, recursive, or dynamical Soft Drop grooming algorithms \cite{Frye:2017yrw,Dreyer:2018tjj,Mehtar-Tani:2019rrk} may potentially address all of these theoretical issues in one fell swoop, but may be more challenging for experimental implementation.  We leave a more detailed study of the vices and virtues of modifications to this SD definition of jet flavour to future work.

Ultimately, we would want to implement this flavour definition into an NNLO prediction matched to a parton shower, just as the BSZ flavour algorithm was used in \cite{Gauld:2020deh,Czakon:2020coa}.  Both the \jade algorithm and SD grooming are natively included in \fastjet \cite{Cacciari:2011ma}, and so can be easily, and universally, implemented in any fixed-order numerical code.  
In this context, it would be important to assess the size of non-perturbative corrections, which are known to be large for \jade, even though our intuition tells us that they should be somewhat reduced in our case, because \jade is used only to recluster the constituents of an existing jet, and the impact that IRC unsafety beyond NNLO has on this type of theoretical predictions.
Further, while we only presented expressions relevant for jets in $e^+e^-$ collisions, the algorithms can be modified for jets in a hadron collider with simple changes of coordinates.  This could then be the first realization of jet predictions on experimentally-preferred flavourless anti-$k_T$ jets \cite{Cacciari:2008gp} with a theoretically-necessary IRC safe flavour definition.  We look forward to exploring the new, flavourful frontiers that this inspires in studying QCD.

%\acknowledgements
\section*{Acknowledgements}
We thank Rhorry Gauld for an inspiring talk at the 2021 CERN workshop on ``Jets and their substructure from LHC data".
AL~is supported by the Department of Energy, Contract DE-AC02-76SF00515. 
The work of SC and SM was supported by the Italian Ministry of Research (MUR) under grant PRIN 20172LNEEZ. 
We also acknowledge support from the Royal Society through the project 580986 ``Resum(e) The Path To Discovery".
SC would like to thank the Institute for Particle Physics Phenomenology (Durham University)
for hospitality during the course of this work.

\appendix

\section{Jet Algorithms for $e^+e^-$ Collisions}
In this appendix, we present the different jet algorithms and the SD grooming algorithm appropriate for jets in $e^+e^-$ collisions.
\subsection{Soft Drop Grooming Algorithm}\label{app:sdalg}

The original algorithm of Ref.~\cite{Larkoski:2014wba} was appropriate for jets produced at a hadron collider, and Ref.~\cite{Frye:2016aiz} modified that definition for jets at an $e^+e^-$ collider.
Given a set of constituents of a jet with radius $R$, the SD grooming algorithm \cite{Larkoski:2014wba} proceeds in the following way:
\begin{enumerate}
\item Recluster the jet with a sequential $k_T$-type \cite{Catani:1991hj,Catani:1993hr,Ellis:1993tq} jet algorithm.  This produces an infrared and collinear (IRC) safe branching history of the jet.  The $k_T$ clustering metric for jets in $e^+e^-$ collisions is
\begin{equation}
d^{e^+e^-}_{ij}=\min\left[
E_i^{2p},E_j^{2p}
\right](1-\cos\theta_{ij})\,,
\end{equation}
where $E_i, E_j$ are the energies of particles $i$ and $j$ and $\theta_{ij}$ is their relative angle.  $p$ is a real number that defines the particular jet algorithm.  The original implementation of SD was restricted to reclustering with the Cambridge/Aachen algorithm ($p=0$) \cite{Dokshitzer:1997in,Wobisch:1998wt,Wobisch:2000dk}.

\item Sequentially step through the branching history of the reclustered jet.  For $e^+e^-$ collisions, we require\footnote{In its implementation in \fastjet, this SD constraint corresponds to setting {\tt SymmetryMeasure} to be {\tt theta\_E}.  Different choices exist for the precise implementation of the SD constraint, but they are all identical in the collinear limit.}
\begin{equation}\label{eq:sdcrit}
\frac{\min[E_i,E_j]}{E_i+E_j}>\zcut \left(
\frac{\theta_{ij}^2}{R^2}
\right)^{\beta}\,.
\end{equation}
If the branching fails this requirement, then the softer of the two daughter branches is removed from the jet.  The SD groomer then continues to the next branching in the remaining clustering history.  

\item The procedure continues until the SD criterion of Eq.~\ref{eq:sdcrit} is satisfied.  At that point, SD terminates, and returns the jet groomed of the branches that failed the SD criterion.

\end{enumerate}

\subsection{Durham (or $k_T$) Algorithm}\label{app:durham}

 The Durham clustering algorithm~\cite{Catani:1991hj}, also known as $k_T$ algorithm, proceeds in the following way:
\begin{enumerate}
    \item For all pairs of (pseudo) particles $i$, $j$ in the event calculate the distance
    \begin{equation}
        y_{ij}=\frac{2\min{[E_i^2, E_j^2]}(1-\cos\theta_{ij})}{Q^2},
    \end{equation}
    where $E_i$ and $E_j$ are the particles' energies, $\theta_{ij}$ is the angle between their three-momenta and $Q$ is the centre-of-mass energy.
    
    \item If all $y_{ij}>y_\text{cut}$, then stop. The number of jets is defined to be equal to the number of pseudoparticles left.
    
    \item Otherwise \emph{recombine} the pair with the smallest value of $y_{ij}$ into a single pseudoparticle according to a particular, recombination scheme (for instance in the so-called $E$-scheme, one sums their four momenta). Go back to step $1$.

\end{enumerate}

\subsection{BSZ (or flavour-$k_T$) Algorithm}\label{app:bsz}

The BSZ or flavour-$k_T$ clustering algorithm~\cite{Banfi:2006hf} closely follows the standard $k_T$ algorithm described above, but it features a flavour-aware distance that ensures IRC safety for flavoured jets. It proceeds as follows
\begin{enumerate}
    \item For all pairs of (pseudo) particles $i$, $j$ in the event calculate the distance
    \begin{equation}
        y^{(F)}_{ij}=\frac{2(1-\cos\theta_{ij})}{Q^2}     \times
   \begin{cases}
    \max{[E_i^2, E_j^2]}^\frac{\alpha}{2}   \min{[E_i^2, E_j^2]}^{1-\frac{\alpha}{2}}, \\    \quad \quad \quad \quad  \quad \quad\quad \text{if the softer of $i,j$} \text{ is flavoured},
  \\
   \min{[E_i^2, E_j^2]}\, , \quad \text{if the softer of $i,j$ is flavourless},
   \end{cases}
    \end{equation}
    where $\alpha$ is a free parameter in the range $0< \alpha\le 2$. The flavour of a pseudoparticle is obtained by simply summing the flavour of its constituents. 
    
    \item If all $  y^{(F)}_{ij}>y_\text{cut}$, then stop. The flavour of each jet is obtained by simply summing the flavour of its constituents
    
    \item Otherwise \emph{recombine} the pair with the smallest value of $y^{(F)}_{ij}$ into a single pseudoparticle according to a particular, recombination scheme (for instance in the so-called $E$-scheme, one sums their four momenta). Go back to step $1$.
    
\end{enumerate}

\section{Calculations in the Triple-Collinear Limit}\label{app:tricoll}

For identification of the divergences that spoil IRC safety of the default SD flavor at NNLO, in Sec.~\ref{sec:nnloircunsafe}, we studied the behavior in the soft and collinear limit of $q\bar q$ emission from a hard quark $Q$.  From the expression in \cite{Catani:1999ss}, the soft and collinear squared matrix element for the process $Q \to Qq\bar q$ is:
\begin{align}
\vert{\cal M}(z_q,z_{\bar q})\vert^2  \propto \frac{z_qz_{\bar q}\left(\theta_{q\bar q}^2(\theta_{Qq}^2+\theta_{Q\bar q}^2)-(\theta_{Qq}^2-\theta_{Q\bar q}^2)^2\right)+\theta_{q\bar q}^2(z_q^2\theta_{Qq}^2+z_{\bar q}^2\theta_{Q\bar q}^2)
}{z_qz_{\bar q}\theta_{q\bar q}^4(z_q+z_{\bar q})^2\left(
z_q\theta_{Q q}^2+z_{\bar q}\theta_{Q\bar q}^2
\right)^2}\,,
\end{align}
in the coordinates introduced in Sec.~\ref{sec:nnloircunsafe}.  Just for isolation of divergences, we ignore overall color and coupling factors, focusing on the kinematic dependence.  The kinematic dependence of the differential soft and collinear three-body phase space $d\Pi_3$ is
\begin{equation}
d\Pi_3\propto  \frac{z_q z_{\bar q}\,dz_q\, dz_{\bar q}\, d\theta_{q\bar q}^2\, d\theta_{Qq}^2\, d\theta_{Q\bar q}^2}{\sqrt{2\theta_{q\bar q}^2\theta_{Qq}^2+2\theta_{q\bar q}^2\theta_{Q\bar q}^2+2\theta_{Qq}^2\theta_{Q\bar q}^2-\theta_{q\bar q}^4-\theta_{Qq}^4-\theta_{Q\bar q}^4}}\,.
\end{equation}
Then, the product of the differential phase space and the squared matrix element is
\begin{align}
d\Pi_3\, \vert{\cal M}(z_q,z_{\bar q})\vert^2&\propto \frac{dz_q\, dz_{\bar q}\, d\theta_{q\bar q}^2\, d\theta_{Qq}^2\, d\theta_{Q\bar q}^2}{\sqrt{2\theta_{q\bar q}^2\theta_{Qq}^2+2\theta_{q\bar q}^2\theta_{Q\bar q}^2+2\theta_{Qq}^2\theta_{Q\bar q}^2-\theta_{q\bar q}^4-\theta_{Qq}^4-\theta_{Q\bar q}^4}} \\
&
\hspace{-1cm}\times\frac{z_qz_{\bar q}\left(\theta_{q\bar q}^2(\theta_{Qq}^2+\theta_{Q\bar q}^2)-(\theta_{Qq}^2-\theta_{Q\bar q}^2)^2\right)+\theta_{q\bar q}^2(z_q^2\theta_{Qq}^2+z_{\bar q}^2\theta_{Q\bar q}^2)
}{\theta_{q\bar q}^4(z_q+z_{\bar q})^2\left(
z_q\theta_{Q q}^2+z_{\bar q}\theta_{Q\bar q}^2
\right)^2}\nonumber\,.
\end{align}

We note that under a homogeneous rescaling of the energy fractions $z_q\to \lambda z_q$ and $z_{\bar q}\to \lambda z_{\bar q}$, the product of the differential phase space and matrix element is invariant:
\begin{align}
\lambda^2d\Pi_3\, \vert{\cal M}(\lambda z_q,\lambda z_{\bar q})\vert^2 = d\Pi_3\, \vert{\cal M}(z_q,z_{\bar q})\vert^2\,.
\end{align}

\section{Numerical Results for a $gg$ Final State}\label{app:num_gg_final}

\begin{figure}[!htb]
\begin{center}
  \includegraphics[width=.49\textwidth]{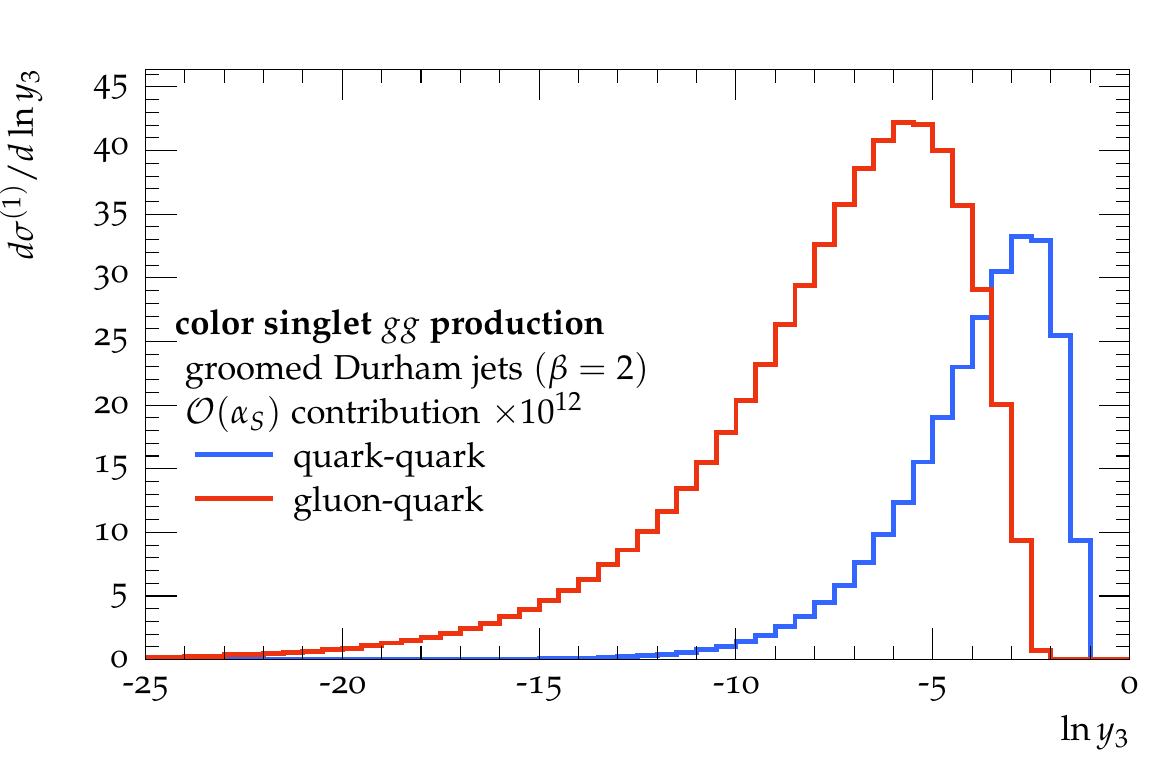}
  \includegraphics[width=.49\textwidth]{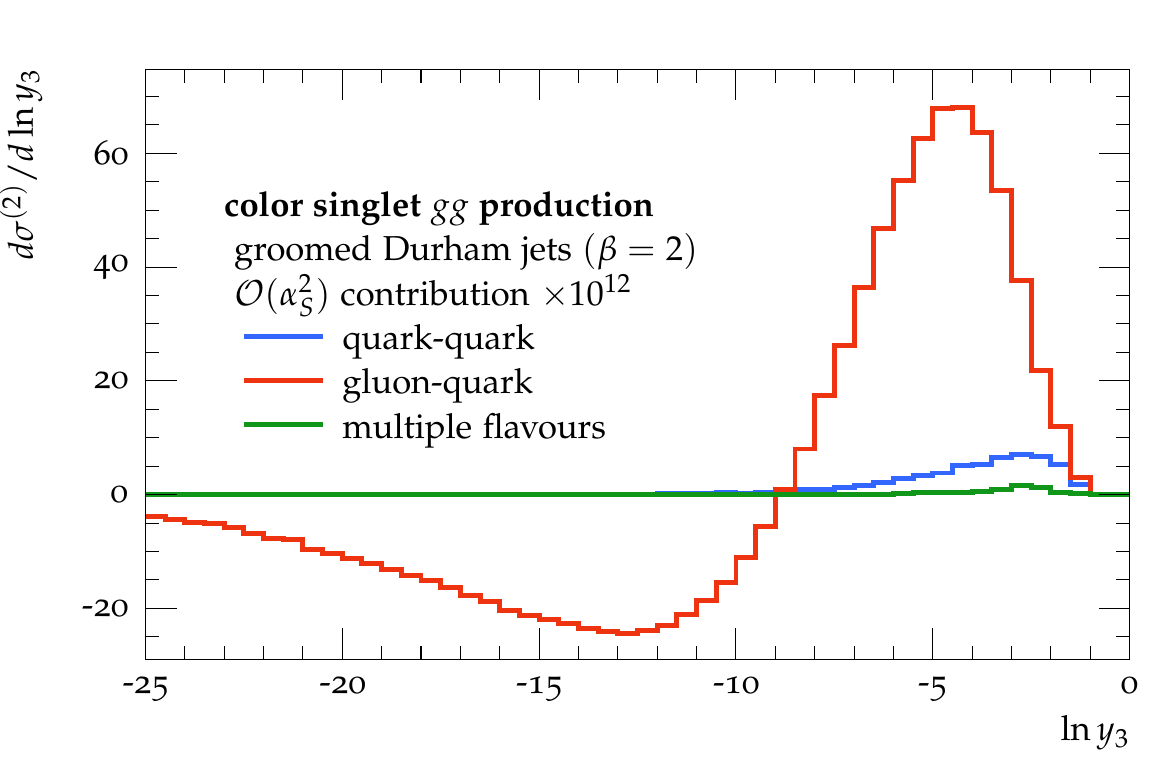}
\caption{NLO (left) and NNLO (right) contributions to the cross section as a function of the $y_3$ jet
  resolution of the event, for different assignments of flavour to the two jets
  obtained from Durham clustering, according to the jet constituents after SD grooming with $\beta = 2$ reclustered with \jade.}\label{numerical_tests_gg}
\end{center}
\end{figure}

Numerical results of the same tests as in
Sec.~\ref{sec:numerical}, but for a final state that at Born level features two gluons
in an overall colour singlet state, can be found in Fig.~\ref{numerical_tests_gg}. Technically, we achieve this by calculating
the process $\mu^+\mu^-\to H\to gg$ in the effective theory where the top loop
inducing the coupling of gluons to the Higgs boson is integrated out. The setup
and tools are otherwise unchanged from Sec.~\ref{sec:numerical}.

\bibliography{jet_flavour}

\end{document}